\documentclass[11pt]{article}
\usepackage[ margin=1in]{geometry}
\usepackage[affil-it]{authblk}
\usepackage{setspace}
\usepackage[sort&compress,numbers]{natbib}
\usepackage[font=small]{caption}
\usepackage{graphicx}% Include figure files
\usepackage{amsmath,amssymb,bm}% bold math
\usepackage{xr}
\usepackage{color}
\usepackage{xcolor}
\usepackage[normalem]{ulem}
\usepackage{wasysym}
\usepackage{lineno}
\usepackage{gensymb}
\usepackage[T1]{fontenc}
\usepackage{booktabs}
\usepackage[section]{placeins}
\usepackage{textcomp}
\usepackage[colorlinks=true, linkcolor=black, urlcolor=blue, citecolor=blue]{hyperref}
\usepackage{verbatim}
\usepackage[utf8]{inputenc}\usepackage[T1]{fontenc}

\makeatletter

\newcommand{\Rmnum}[1]{\expandafter\@slowromancap\romannumeral #1@}

\newcommand{\wm}{\omega_m}
\newcommand{\w}{\omega}

\newcommand{\V}{{v_b}}

\makeatother

\begin{document}
\title{Modulation of quantum geometry and its coupling to pseudo-electric field by dynamic strain}

\author[1$\dagger$]{Surat Layek}

\author[1]{Mahesh A. Hingankar}
\affil[1]{Department of Condensed Matter Physics and Materials Science, Tata Institute of Fundamental Research, Homi Bhabha Road, Mumbai 400005, India.}

\author[1]{Ayshi Mukherjee}
\author[2]{Atasi Chakraborty}
\affil[2]{Institut f\"{u}r Physik, Johannes Gutenberg Universit\"{a}t Mainz, D-55099 Mainz, Germany.}

\author[1]{Digambar A. Jangade}
\author[1]{Anil Kumar}
\author[1]{L. D. Varma Sangani}
\author[1]{Amit Basu}
\author[3,4]{R Bhuvaneswari}
\affil[3]{Theoretical Sciences Unit, Jawaharlal Nehru Centre for Advanced Scientific Research, Bangalore 560064, India.}
\affil[4]{School of Electrical and Electronics Engineering, SASTRA Deemed University, Thanjavur 613401, India.}

\author[5]{Kenji Watanabe}
\affil[5]{Research Center for Functional Materials, National Institute for Materials Science, 1-1 Namiki, Tsukuba 305-0044, Japan.}

\author[6]{Takashi Taniguchi}
\affil[6]{International Center for Materials Nanoarchitectonics, National Institute for Materials Science,  1-1 Namiki, Tsukuba 305-0044, Japan.}

\author[7]{Amit Agarwal}
\affil[7]{Department of Physics, Indian Institute of Technology, Kanpur 208016, India.}

\author[3$\ddagger$]{Umesh V. Waghmare}

\author[1*]{Mandar M. Deshmukh}
\affil[$\dagger$]{\textnormal{suratlayek91@gmail.com}}
\affil[$\ddagger$]{\textnormal{waghmare@jncasr.ac.in}}
\affil[*]{\textnormal{deshmukh@tifr.res.in}}

\date{}
\maketitle
\newpage
%\linenumbers
\setlength{\columnsep}{5cm}

\section*{Abstract}
Two-dimensional materials are a fertile ground for exploring quantum geometric phenomena, with Berry curvature and its first moment, the Berry curvature dipole, playing a central role in their electronic response. These geometric properties influence electronic transport and result in the anomalous and nonlinear Hall effects, and are typically controlled using static electric fields or strain. However, the possibility of modulating quantum geometric quantities in real-time remains unexplored. Here, we demonstrate the dynamic modulation of Berry curvature and its moments, as well as the generation of a pseudo-electric field and their coupling. By placing heterostructures on a membrane, we introduce oscillatory strain together with an in-plane AC electric field and measure Hall signals that are modulated at linear combinations of the frequencies of strain and electric field. We also present direct experimental and theoretical evidence for coupling between pseudo-electric field and quantum geometry that results in an unusual dynamic strain-induced Hall response. This approach opens up a new pathway for controlling quantum geometry on demand, moving beyond conventional static perturbations. The coupling of pseudo-electric field with Berry curvature provides a framework for external electric field-free anomalous Hall response and opens new avenues for probing the topological properties.

\section*{Introduction}

Two-dimensional (2D) materials have emerged as an exciting platform in condensed matter physics and material sciences, offering unique opportunities to study, tune, and probe fundamental quantum phenomena.
From graphene to transition metal dichalcogenides (TMDC) and onwards to moir\'e superlattices, these atomically thin systems exhibit properties such as tunable band gaps~\cite{ohta_controlling_2006,zhang_direct_BLG_2009}, high carrier mobility, strong spin-orbit coupling~\cite{arp_intervalley_2024}, and pronounced excitonic effects, making them ideal candidates for next-generation quantum devices.
Among the various approaches to tailor their properties, the application of mechanical strain stands out as a versatile and non-invasive tool.
Strain enables modification of band structures, valley polarization~\cite{li_valley_2020,qi_strain-induced_2024}, and induces phase transitions~\cite{cenker_reversible_2022,nicholson_uniaxial_2021}, thereby significantly altering the electronic and optical responses.

A particularly intriguing aspect of 2D materials is the role of quantum geometry in governing their physical responses~\cite{jiang_revealing_2025}, mainly encoded by Berry curvature and the Berry curvature dipole (BCD)~\cite{kruchkov_quantum_2022,liu_quantum_2024,adak_tunable_2024}.
Berry curvature acts as an effective valley contrasting magnetic field in momentum space~\cite{xiao_valley-contrasting_2007}, influencing the anomalous Hall~\cite{nagaosa_anomalous_2010} and valley Hall~\cite{mak_valley_2014,shimazaki_generation_2015,sui_gate-tunable_2015} effects in systems where time-reversal or inversion symmetry is broken. 
In contrast, the BCD, which captures the asymmetric distribution of Berry curvature, generates the nonlinear Hall effect (NLHE) and dominates the Hall transport in time-reversal invariant systems with broken crystalline rotation symmetry.
Together, they offer a comprehensive framework to understand and predict the transport behavior of 2D materials.
Strain introduces an additional degree of freedom in this framework to manipulate both Berry curvature and BCD.
By modifying lattice symmetries and electronic wavefunctions, strain can modify Berry curvature distribution in momentum space and induce asymmetries that significantly alter the BCD~\cite{hu_nonlinear_2022}. 

Existing studies have explored how static strain affects Berry curvature and the BCD in 2D materials.
In TMDCs, strain has been shown to enhance valley-contrasting phenomena, such as the valley Hall effect~\cite{zhao_enhanced_2020,shayeganfar_strain_2022,son_strain_2019}.
Moir\'e systems such as twisted bilayer graphene exhibit strain-tunable Berry curvature distributions, which can modify band topology by altering Chern numbers, leading to nontrivial topological transport phenomena~\cite{zhang_giant_2022,bi_designing_2019,pantaleon_tunable_2021}.
Experimental and theoretical studies have demonstrated that strain-induced symmetry breaking enhances the BCD, giving rise to a pronounced nonlinear Hall effect in materials like WTe$_\text{2}$ and MoTe$_\text{2}$~\cite{ma_observation_2019, kang_nonlinear_2019}. 
Recent studies on twisted double bilayer graphene (TDBG) have demonstrated that the BCD can sense topological transitions in moir\'e superlattices, underscoring the role of strain in tuning Berry curvature distribution and topological properties~\cite{sinha_berry_2022,chakraborty_nonlinear_topological_2022}.
Although these findings highlight the sensitivity of Berry curvature and BCD to strain, they predominantly address static strain, which is challenging to control due to the inherent unpredictability of stacking-induced strain. 

\begin{figure*}
    \centering
    \includegraphics[width=16.5cm]{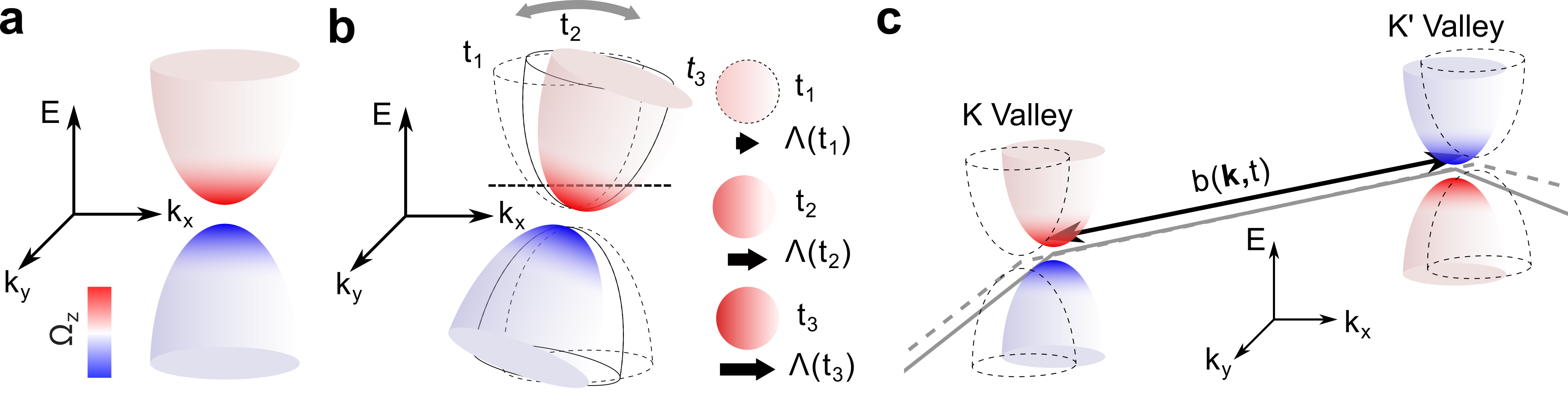}
    \caption{ \label{fig:fig0_modulation_intro}
    \textbf{Schematic illustration of time-dependent strain-induced modulation of quantum geometry and pseudo-electric field generation.} \textbf{a,} Parabolic  bands with Berry curvature ($\Omega_z$) distribution indicated by the red–blue colormap. 
    \textbf{b,} Time-dependent strain tilts the parabolic bands periodically, leading to a modulation of the Berry curvature dipole (BCD, $\Lambda(t)$) over time. The right panels show the instantaneous Berry curvature distributions at different times within one strain cycle, with black arrows indicating the magnitude of the BCD.
    \textbf{c,} Strain also modulates the momentum-space separation b(\textbf{k},t) between the two valleys over time. This time-dependent valley displacement produces a pseudo-electric field.
    }
\end{figure*}

In this work, we introduce adiabatic oscillatory strain as a tool to dynamically modulate Berry curvature and BCD in 2D materials, providing direct access to their temporal evolution~\cite{r_sensing_2024,chen_crossed_2024}.
Figure~\ref{fig:fig0_modulation_intro}a illustrates a representative untilted parabolic band with a symmetric Berry curvature distribution over the Fermi surface--corresponding to a vanishing BCD.
Dynamic strain perturbs the symmetry in two distinct ways--band tilting (Figure~\ref{fig:fig0_modulation_intro}b) and valley separation (Figure~\ref{fig:extended_fig_0}c), each leaving a unique fingerprint in transport. 
The oscillatory strain periodically tilts the electronic bands, distorting the Berry curvature distribution over the Fermi surface, leading to BCD modulation as schematically shown in figure~\ref{fig:fig0_modulation_intro}b.
Also, the strain modulates the separation between the $K$ and $K'$ valleys as shown in figure~\ref{fig:fig0_modulation_intro}c.
Microscopically, this modifies the hopping amplitudes in the hexagonal lattice, inducing a time-dependent gauge field~\cite{vozmediano_gauge_2010,ramezani_pseudo_2013} that leads to pseudo-electric field~\cite{ilan_pseudo-electromagnetic_2020,zhao_acoustically_2022,hadadi_pseudo_2023,sela_quantum_2020}(See Supplementary Information section~III).

This capability allows for real-time control of Berry curvature and helps disentangle other nonlinear effects from quantum geometric contributions.
To explore strain-induced modulation of quantum geometry, we focus on TDBG and Bernal Bilayer graphene (BLG) for their complementary features.
TDBG is a highly tunable system where the twist angle between two BLG sheets creates flat electronic bands~\cite{bistritzer_moire_2011,koshino_band_2019}.
The tunability of its band structure via perpendicular electric field and carrier density, combined with Berry curvature hotspots near the band edges~\cite{adak_tunable_2020,haddadi_moire_2020,sinha_bulk_valley_2020}, makes it ideal for studying strain-driven Berry curvature dynamics.
BLG, in contrast, serves as a simpler and well-understood reference system, where a perpendicular electric field breaks inversion symmetry and opens a bandgap at the charge neutrality point (CNP)~\cite{zhang_direct_BLG_2009}.
To enable controlled strain application in these systems, we fabricate our devices on trampoline-like, freely suspended silicon nitride membranes.
This platform allows mechanical coupling without substrate interference and is particularly well suited for delicate moir\'e heterostructures.
Through experiments on TDBG and BLG devices, we show that strain oscillations modify the nonlinear Hall signal--arising from the time-dependent BCD--at specific mixed frequencies, directly revealing the dynamic response of quantum geometry.
Alongside, we report direct observation of pseudo-electric field coupling to the Berry curvature, resulting in an external field-free anomalous Hall voltage.
Our findings, using versatile 2D systems, introduce a novel approach for dynamic tuning of quantum geometric properties of materials of any dimensionality beyond static perturbations.

\section*{Experimental Results:}

\begin{figure*}
    \centering
    \includegraphics[width=16cm]{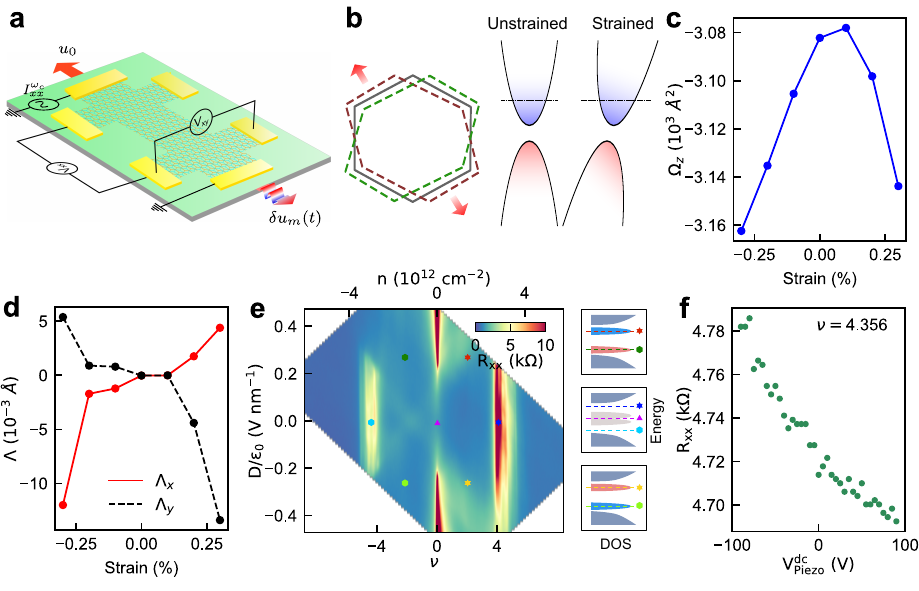}
    \caption{ \label{fig:fig1_static_strain}\textbf{Tailoring of band structure and transport properties – Berry curvature, BCD in TDBG induced by controlled static strain.}  
\textbf{a,}~Schematic of the device on a silicon nitride membrane, where static strain $u_0$ and oscillatory strain $\delta u_m$ are independently applied via two piezoelectric stacks on a strain cell. Hall voltages ($V_{xy}^\omega$) are measured at characteristic mixed frequencies. 
\textbf{b,}~Illustration of a honeycomb lattice under uniaxial strain, which breaks the $C_3$ symmetry. Gray indicates the unstrained lattice, while brown and green represent lattice under tensile and compressive strain, respectively. The right panel shows the tilted band structure under strain with a colormap denoting Berry curvature and a dashed line marking the Fermi level. 
\textbf{c,}~Theoretically calculated Berry curvature at a $k$ point near the valence band maxima in the $K$ valley of TDBG under uniaxial strain.
\textbf{d,}~Strain dependence of the $x$ and $y$ components of the Berry curvature dipole ($\Lambda_x$, $\Lambda_y$) at the valence band maxima in the $K$ valley.
\textbf{e,}~Measured longitudinal resistance $R_{xx}$ is plotted as a function of carrier density $n$ (top $x$-axis), filling factor $\nu$ (bottom $x$-axis) and perpendicular electric field $D/\epsilon_0$ ($y$-axis), at $T = 1.5~\text{K}$. The two high-resistance peaks correspond to full filling ($\nu = \pm 4$) of the flat bands. 
The schematic insets on the right depict the corresponding density of states (DOS) configurations, where colored markers indicate the Fermi level positions at selected points in the phase diagram.
Red and blue colors represent Berry curvature of opposite sign in the flat bands, which reverse upon flipping the the perpendicular electric field.
\textbf{f,}~Variation in $R_{xx}$ with DC piezo voltage ($V_{\text{piezo}}^{dc}$) at $n = 3.46\times10^{12}\mathrm{cm^{-2}}$ ($\nu = 4.356$), $D/\epsilon_0 = 0~\text{V/nm}$, and $T = 1.5~\text{K}$ confirms strain application. Negative and positive $V_{\text{piezo}}^{dc}$ correspond to compressive and tensile strain, respectively.
}
\end{figure*}

The TDBG device, consisting of two bilayer graphene sheets stacked with a twist angle of 1.1$^\circ$, and the BLG device are fabricated on 300~nm thick and $300~\mathrm{\mu m} \times 300~\mathrm{\mu m}$ silicon nitride (SiN) membranes to facilitate the strain application~\cite{hicks_piezoelectric-based_2014,liu_continuously_2024} (see Supplementary Figure~S4 and Supplementary Information section~VII and VIII).
Finite element analysis COMSOL simulation is used for placement of the device stack on the SiN membrane for optimal strain application (see Supplementary Information section~IX).
The devices are configured in a dual-gated geometry, enabling independent control of the perpendicular electric field~$D/\epsilon_0$ and carrier density~$n$ (see Supplementary Information section~X.1 for details).
Time-dependent strain~$\delta u_m(t)$ and DC strain~$u_0$ are applied to the device (see Figure~\ref{fig:fig1_static_strain}a) using a Razorbill CS130 strain cell, and all measurements are performed at 1.5 K (further details on measurements are provided in Supplementary Information section~X).

A sinusoidal strain of frequency $\w_m$,~$\delta u_m(t) \propto \text{sin}(\omega_m t)$ is applied to the devices via the strain cell.
$\w_m$ is kept low to ensure adiabatic evolution of the electronic wavefunction.
In addition to the strain modulation, an AC current $I_{xx}^{\omega_c}$ at frequency $\omega_c$ is applied to the device and the resulting Hall response $V_{xy}^\omega$ is measured at mixed frequencies, $\omega = \omega_m \pm \alpha\omega_c$ ($\alpha = 0,1,2$).
Phase-sensitive detection of the Hall response is performed by separately analyzing the real and imaginary components of the AC signal, depending on the frequency combination.
In the following sections, we present the experimental results and analyze the corresponding Hall responses.

Figure~\ref{fig:fig1_static_strain}b depicts the honeycomb lattice under uniaxial strain.
The unstrained lattice is shown in gray, while the strained lattice under tension and compression is represented by brown and green, respectively.
Strain in graphene can tilt the band structure~\cite{pereira_tight-binding_2009}, altering the Berry curvature distribution over the Fermi surface, as schematically depicted. 
Consequently, the valley-specific Berry curvature and BCD are modified. 
The theoretically calculated variation of Berry curvature and BCD with strain within $K$-valley for 1.1$^\circ$ TDBG  is shown in Figure~\ref{fig:fig1_static_strain}c and \ref{fig:fig1_static_strain}d, respectively.
The Berry curvature and BCD are calculated at the maxima of the valence band (see Supplementary Information section~I), considering uniaxial strain along the zig-zag direction of graphene.
Both theoretical plots emphasize the sensitivity of Berry curvature and BCD to strain in TDBG.

Figure~\ref{fig:fig1_static_strain}e presents a colormap of the longitudinal resistance $R_{xx}$ as a function of the filling factor ($\nu$, bottom x-axis), carrier density ($n$, top x-axis), and perpendicular electric field ($D/\epsilon_0$, y-axis) at 1.5 K under zero strain.
At $D/\epsilon_0 = 0$, the two flat bands in TDBG remain gapless; however, as a perpendicular electric field is applied, a bandgap opens between the two flat bands, which is observed as a prominent resistance peak at $\nu = 0$.
In addition, well-defined resistance peaks appear at $\nu = \pm 4$, corresponding to the full filling of the moir\'e flat bands; this measurement highlights the tunability of the band structure of TDBG by a perpendicular electric field, consistent with prior studies on TDBG~\cite{koshino_band_2019,shen_correlated_2020,adak_tunable_2020,sinha_bulk_valley_2020}.

The longitudinal resistance ($R_{xx}$) changes notably with applied DC strain. 
Figure~\ref{fig:fig1_static_strain}f shows $R_{xx}$ as a function of the DC piezo voltage ($V_{\text{piezo}}^{dc}$), which controls the applied strain. 
The resistance, measured at a fixed carrier density ($n = 3.46\times 10^{12}~\mathrm{cm^{-2}}$) slightly above full filling ($\nu = 4$) at 1.5 K, responds to both tensile ($V_{\text{piezo}}^{dc} > 0$) and compressive ($V_{\text{piezo}}^{dc} < 0$) strain. 
In 2D materials, strain modifies interatomic distances and breaks lattice symmetries, leading to significant changes in carrier mobility~\cite{datye_strain-enhanced_2022}, band structure, and Berry curvature distribution~\cite{hou_strain_2024}.
The observed strain dependence of $R_{xx}$ thus confirms effective strain application and underscores its impact on the electronic properties of the device.
\subsection*{Berry Curvature Dipole Modulation}
A key outcome of strain modulation in TDBG is its profound impact on the BCD ($\vec\Lambda$) and its role in non-linear Hall transport~\cite{zhang_giant_2020,sinha_berry_2022}.
In 2D devices such as TDBG and BLG, a pre-existing static strain ($u_0$) often arises due to fabrication-induced imperfections or can be deliberately applied via a DC voltage to the strain cell. 
This static strain breaks the $C_3$ symmetry of the lattice, leading to an asymmetric Berry curvature distribution in momentum ($k$) space. 
Such symmetry breaking is responsible for a non-zero BCD, which generates nonlinear Hall voltages (Supplementary Information section~XI), in presence of an in-plane electric field $\vec{E}_{\omega_c}=\hat{x} E_0 \sin(\omega_c t)$,
\begin{equation}\label{eqn:NLH}
    \vec{j^{(2)}}=\frac{e^3\tau}{2(1-i\omega_c\tau)}\hat{z}\times\vec{E}_{\omega_c}(\vec{\Lambda}\cdot\vec{E}_{\omega_c}).
\end{equation}
In addition to the static strain, when an oscillatory strain $\delta u_m(t)\propto \text{sin}(\omega_m t)$, experimentally controlled by an AC piezoelectric voltage, $V_{\omega_m} = V_m~\text{sin}(\omega_m t)$ is applied to TDBG, the non-linear Hall current exhibits a strain-dependent variation, as the applied mechanical deformation introduces a dynamic component to the BCD, given by $\vec\Lambda(u_0 + \delta u_m) = \vec\Lambda_0 + \vec{\delta \Lambda_m}~\text{cos}(\omega_m t)+ \dots$.
This manifests as mixing of frequencies between the applied in-plane electric field and the oscillatory strain, leading to a non-linear Hall response that displays sidebands at combined frequencies of the strain ($\omega_m$) and current modulation ($\omega_c$), 
\begin{equation}\label{eqn:NLH_mixed_theory}
 j_y^{(2)} \propto \frac{E_0^2 \Lambda_0}{2} - \frac{E_0^2 \Lambda_0}{2} \text{cos}(2\omega_c t) + \frac{E_0^2 \delta \Lambda_m}{2} \text{cos}(\omega_m t) - \frac{E_0^2 \delta \Lambda_m}{4} \left[\text{cos}((\omega_m + 2\omega_c)t) + \text{cos}((\omega_m - 2\omega_c)t)\right].
\end{equation}
Here,
\begin{equation}\label{eqn:del_lambda}
\delta \Lambda_m = \left(\frac{\partial \Lambda}{\partial u}\right)_{u_0} \left(\frac{\partial u}{\partial V}\right)_{V_0} V_m~\omega_m \delta t
\end{equation}
is the change of BCD due to strain over a small time interval $\delta t$, and $\Lambda_0 = \Lambda(u_0)$ is the static BCD. $\left(\frac{\partial \Lambda}{\partial u}\right)_{u_0}$ captures the dependence of the BCD on strain, determined by the band structure, while $\left(\frac{\partial u}{\partial V}\right)$ depends on the experimental setup and the effective coupling between piezo voltage and strain (see Supplementary Information section~III for detailed calculation).

The time-independent component of the BCD ($\Lambda_0$) generates a DC rectification and a conventional nonlinear Hall voltage at $2\omega_c$.
The oscillatory component of the BCD ($\delta \Lambda_m$) contributes a rectification at $\omega_m$ and produces sidebands at $\omega_m \pm 2\omega_c$.
Since these frequency components arise from dynamic strain-induced BCD oscillations, they exhibit phase differences relative to the applied strain and current. 
Accurate detection of these nonlinear Hall components requires phase-sensitive measurements.
The nonlinear Hall signals appear 90$^\circ$ out of phase with the applied alternating current, requiring measurement of its imaginary component to capture contributions at $2\omega_c$, $\omega_m$, $\omega_m + 2\omega_c$, and $\omega_m - 2\omega_c$ (for details see Supplementary Information section~III).

\begin{figure*}
    \centering
    \includegraphics[width=16cm]{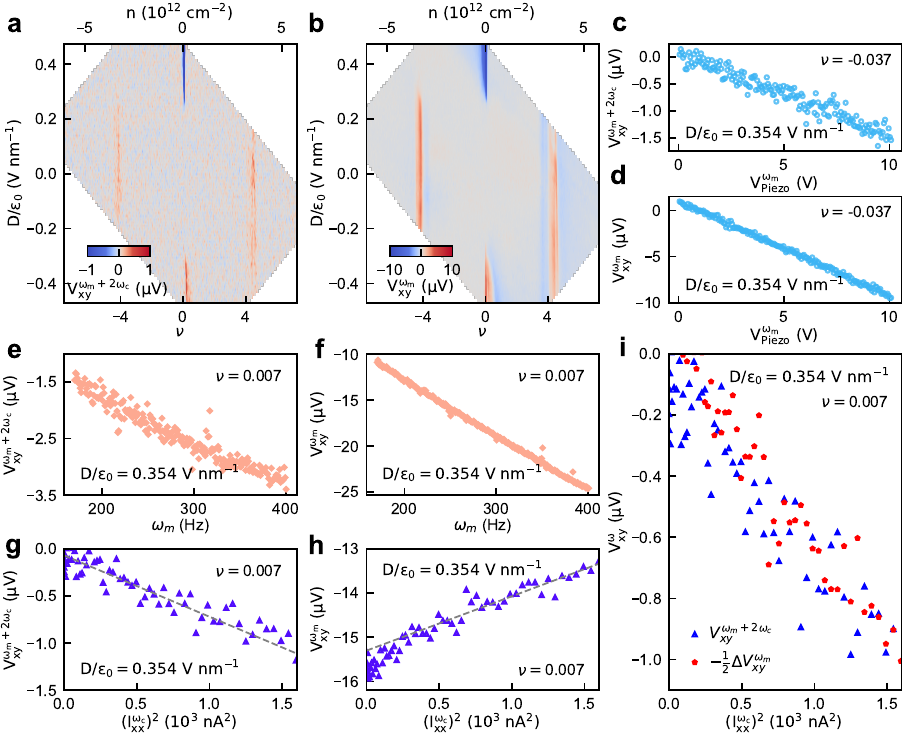}
    \caption{ \label{fig:fig2_BCD_modulation_TDBG}\textbf{Nonlinear Hall signals encode the modulation of BCD and pseudo-electric field due to the application of dynamic strain.}
\textbf{a, b,}~ Nonlinear Hall voltages at $\omega_m + 2\omega_c$ ($V_{xy}^{\omega_m + 2\omega_c}$) (\textbf{a}) and $\omega_m$ ($V_{xy}^{\omega_m}$) (\textbf{b}) as functions of carrier density ($n$) and perpendicular electric field ($D/\epsilon_0$) at $T = 1.5 \, \text{K}$. Measurements were performed using a current of 98 nA at $\omega_c = 17$ Hz and a piezoelectric excitation at $\omega_m = 177$ Hz with a peak-to-peak voltage of 10 V. Both signals exhibit antisymmetric behavior under $D$-field reversal, consistent with the sign-change of the BCD in ABAB TDBG.
\textbf{c, d,}~Linear scaling of the Hall voltages at $\omega_m + 2\omega_c$ (\textbf{c}) and $\omega_m$ (\textbf{d}) with the amplitude of strain modulation, represented by the peak-to-peak piezoelectric excitation voltage ($V_{\text{piezo}}^{\omega_m}$), confirming that the BCD modulation is directly proportional to strain modulation amplitude. 
\textbf{e, f,}~Frequency dependence of the nonlinear Hall signals shows linear scaling with $\omega_m$, indicating that the signals are directly proportional to the rate of change of BCD.
\textbf{g, h,}~Linear scaling of the Hall voltages at $\omega_m + 2\omega_c$ (\textbf{g}) and $\omega_m$ (\textbf{h}) with the square of the current amplitude, $(I_{xx}^{\omega_c})^2$, indicates a second-order nonlinear Hall response with respect to the in-plane electric field. Gray dashed lines represent linear fits to the experimental data, confirming the expected quadratic dependence on $I_{xx}^{\omega_c}$.
Notably, $V_{xy}^{\omega_m}$ remains finite even at zero current, an indication to the presence of pseudo-electric field. The residual finite offset at $I_{xx}^{\omega_c}=0$ is arising from a pseudo-electric field generated by time-dependent strain.
\textbf{i,} The extracted BCD-driven component of the nonlinear Hall signal, defined as $\Delta V_{xy}^{\omega_m} = V_{xy}^{\omega_m} - V_{xy}^{\omega_m}(I_{xx}^{\omega_c} = 0)$ from panel (h), is plotted as $- \frac{1}{2} \Delta V_{xy}^{\omega_m}$ (red pentagons) against $(I_{xx}^{\omega_c})^2$. It closely overlaps with the independently measured $V_{xy}^{\omega_m + 2\omega_c}$ signal (blue triangles), confirming the strain-modulated BCD origin of the response.
}
\end{figure*}

Figure \ref{fig:fig2_BCD_modulation_TDBG}a and \ref{fig:fig2_BCD_modulation_TDBG}b present the nonlinear Hall signals at the frequencies $\omega_m + 2\omega_c$ (see Supplementary Information section~XII for $\omega_m - 2\omega_c$ signal) and $\omega_m$, respectively, as functions of carrier density ($n$) or filling factor ($\nu$) and the perpendicular electric field ($D / \epsilon_0$) at 1.5 K under zero magnetic field.
These measurements are performed using an alternating current of 98 nA and a sinusoidal strain generated by a 10~V peak-to-peak AC piezo excitation.
Both the Hall signals at $\omega_m$ and $\omega_m+2\omega_c$ frequencies exhibit antisymmetric behavior under reversal of the perpendicular electric field ($D/\epsilon_0$) near the CNP.
For positive $D$, the Hall signals are negative for $\nu < 0$ and become positive when $D$ is flipped to negative values~(see Supplementary Information section~XIV for data from another TDBG device).
This behavior is consistent with the fact that the Berry curvature and BCD reverse sign with the direction of the electric field for ABAB TDBG~\cite{layek_quantum_2025}.

To confirm that oscillating strain underlies the nonlinear Hall signals, we systematically study their dependence on parameters characterizing the strain, the amplitude and frequency of the strain modulation.
Figure \ref{fig:fig2_BCD_modulation_TDBG}c and \ref{fig:fig2_BCD_modulation_TDBG}e illustrate the scaling of the nonlinear Hall signals at $\omega_m + 2\omega_c$ and $\omega_m$, respectively; the amplitude of the strain modulation, represented by the AC piezo excitation voltage ($V_{\text{piezo}}^{\omega_m}$) at $\nu = -0.037$ and $D/\epsilon_0 = 0.354~\mathrm{V/nm}$ at 1.5 K.
Both $V_{xy}^{\w_m+2\w_c}$ and $V_{xy}^{\wm}$ signals vary linearly with the strain amplitude.
Additionally, as the frequency determines the time derivative of the BCD, we examined the frequency dependence of the Hall signals.
Figure~\ref{fig:fig2_BCD_modulation_TDBG}e and~\ref{fig:fig2_BCD_modulation_TDBG}f illustrate the linear dependence for both $V_{xy}^{\omega_m + 2\omega_c}$ and $V_{xy}^{\omega_m}$ on $\omega_m$.
These results are in agreement with equation (\ref{eqn:NLH_mixed_theory}) and (\ref{eqn:del_lambda}), and offer compelling experimental evidence for the dynamic modulation of the BCD under strain.
Figure~\ref{fig:fig2_BCD_modulation_TDBG}g and~\ref{fig:fig2_BCD_modulation_TDBG}h further depict the scaling of the nonlinear Hall signals at $\omega_m + 2\omega_c$ and $\omega_m$ with the square of the applied current $I_{xx}^{\omega_c}$, at $\nu = 0.007$, $D/\epsilon_0 = 0.354~\mathrm{V/nm}$.
The quadratic current dependence confirms that the nonlinear response stems from a second-order process with the in-plane electric field.
Notably, $V_{xy}^{\omega_m}$ remains finite even at zero current, suggesting contributions beyond simple BCD modulation.
\subsection*{Coupling of Pseudo-Electric Field with Static Berry Curvature}
As outlined earlier, oscillatory strain can also generate a pseudo-electric field via a time-dependent gauge potential.
Unlike real electric fields, the pseudo-electric field $\vec{E}_{\omega_m}$ is valley-dependent and can be written as $\vec{E}_{\omega_m} = \chi \vec{E}_m(t)$~\cite{ilan_pseudo-electromagnetic_2020,zhao_acoustically_2022, hadadi_pseudo_2023}, where $\chi = \pm 1$ denotes the valley chirality.
In the presence of nonzero Berry curvature $\vec{\Omega}$, the semiclassical equation of motion yields an anomalous velocity $e (\vec{\Omega} \times \chi \vec{E}_m)$, that is even under time reversal symmetry.
This chirality-dependent coupling drives carriers in both valleys in the same direction (Extended Figure~\ref{fig:extended_fig_1}a), resulting in an anomalous Hall effect at frequency $\omega_m$ solely dependent on the pseudo-electric field (see row~2 of Table~5 in Supplementary Information section~V), a notable aspect of our work.
While earlier studies have explored couplings between pseudo-electric and pseudo-magnetic fields\cite{zhao_acoustically_2022}, we emphasize that the effect reported here originates from Berry curvature. This mechanism is distinct from pseudo-magnetic fields, which instead arise from spatial strain gradients.

This mechanism also accounts for the deviation from the expected ratio between the signals $V_{xy}^{\omega_m}$ and $V_{xy}^{\omega_m + 2\omega_c}$ observed in Figure~\ref{fig:fig2_BCD_modulation_TDBG}(a–d).
Specifically, the pseudo-electric field contributes independently of the strain-induced modulation of the BCD, leading to a finite value of $V_{xy}^{\omega_m}$ even at $I_{xx}^{\omega_c} = 0$ (Extended Figure~\ref{fig:extended_fig_1}b).
This offset reflects the intrinsic Hall voltage originating completely from the pseudo-electric field.
%Figure~\ref{fig:fig2_BCD_modulation_TDBG}i presents $\Delta V_{xy}^{\omega_m} = V_{xy}^{\omega_m} - V_{xy}^{\omega_m}(I_{xx}^{\omega_c} = 0)$, which represents the BCD-modulated component of the nonlinear Hall voltage.
Figure~\ref{fig:fig2_BCD_modulation_TDBG}i shows the BCD-modulated component of the nonlinear Hall voltage $\Delta V_{xy}^{\omega_m} = V_{xy}^{\omega_m} - V_{xy}^{\omega_m}(I_{xx}^{\omega_c} = 0)$.
As suggested by equation~(\ref{eqn:NLH_mixed_theory}), this term is expected to be twice the magnitude of $V_{xy}^{\omega_m + 2\omega_c}$ with opposite sign. Accordingly, we scale $\Delta V_{xy}^{\omega_m}$ by a factor of $-1/2$, and observe that $- \frac{1}{2} \Delta V_{xy}^{\omega_m}$ accurately overlays with $V_{xy}^{\omega_m + 2\omega_c}$ in Figure~\ref{fig:fig2_BCD_modulation_TDBG}i, in excellent agreement with theoretical expectations.

These observations not only validate the dynamic modulation of the BCD under strain, but also suggest the presence of additional nonlinear transport phenomena beyond the BCD framework.
In particular, the finite offset in $V_{xy}^{\omega_m}$ points toward an intriguing microscopic mechanism--namely, the emergence of strain-induced pseudo-electric field.
If such a pseudo-field is indeed present, it is expected to give rise to Hall-like signals at mixed frequencies $\omega_m\pm\omega_c$ as a consequence (see row~6,7 of table~5 in Supplementary Information section~V).
This motivates our following investigation of the  $V_{xy}^{\omega_m\pm\omega_c}$ response in detail.

\subsection*{Coupling of Pseudo-Electric Field and Band Velocity}

\begin{figure*}
    \centering
    \includegraphics[width=16cm]{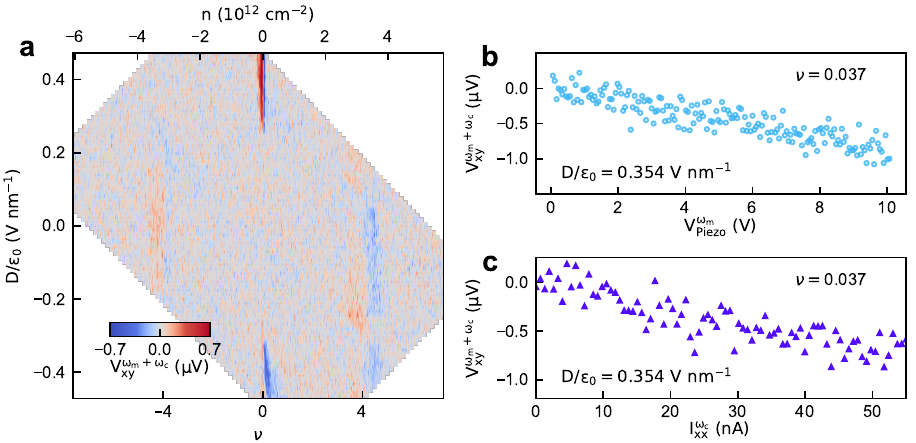}
    \caption{ \label{fig:fig3_BC_modulation_TDBG}\textbf{Band velocity-driven transverse signals show the evidence of a chirality-dependent pseudo-electric field.}
\textbf{a,}~Measured transverse voltage at the mixed frequency $\omega_m + \omega_c$ in TDBG as a function of the perpendicular electric field ($D/\epsilon_0$) and filling factor ($\nu$) at $T = 1.5 \, \text{K}$.
The signal remains positive just below full-band fillings ($\nu \lessapprox -4, 0, +4$) and negative just above ($\nu \gtrapprox -4, 0, +4$), independent of $D/\epsilon_0$. The observed insensitivity to electric field direction indicates a distinct origin from a band velocity–driven mechanism induced by a valley-contrasting pseudo-electric field.
\textbf{b,}~Linear scaling of the Hall-like voltage at $\omega_m + \omega_c$ with the amplitude of strain modulation, represented by the peak-to-peak piezoelectric excitation voltage ($V_{\text{piezo}}^{\omega_m}$).  
\textbf{c,}~Linear scaling of the Hall-like voltage ($V_{xy}^{\omega_m+\omega_c}$) with the amplitude of the applied current ($I_{xx}^{\omega_c}$), confirming the linear dependence on the in-plane electric field.
}
\end{figure*}
Having established the presence of a finite $ V_{xy}^{\omega_m} $ offset, indicative of a pseudo-electric field, we now examine its dynamical consequences--specifically, the appearance of transverse voltages at mixed frequencies $ \omega_m \pm \omega_c $.
These components are not captured within the conventional BCD framework but emerge naturally from the interplay between a strain-induced pseudo-electric field and the applied current.
Time-dependent strain gives rise to a gauge potential $ \chi\vec{A}_{\omega_m} $~\cite{levy_strain-induced_2010,vozmediano_gauge_2010,ramezani_pseudo_2013}, whose temporal derivative defines the pseudo-electric field, $\vec{E}_{\omega_m}(t) = -\chi\partial_t \vec{A}_{\omega_m}(t)$.

Within the Boltzmann transport formalism (Supplementary Section~V), this pseudo-electric field, when combined with an in-plane AC current at frequency $\omega_c$, leads to a second-order nonlinear response at $\omega_m \pm \omega_c$. 
Notably, the corresponding current arises not from the anomalous velocity, which is Berry curvature dependent, but from the band velocity~($\vec\V=\frac{\partial\varepsilon}{\partial\vec k}$) term, yielding a transverse voltage that is Hall-like in appearance but distinct in origin,
\begin{equation*}
\vec{j}^{\omega_m + \omega_c} \propto -e \chi f_{2}^{\omega_m, \omega_c} \vec{\V}-e \chi f_{2}^{\omega_c, \omega_m} \vec{\V}~;\quad f_2^{\omega_j,\omega_k} = \frac{(e\tau)^2}{4[1 + i(\omega_j + \omega_k)\tau](1 + i\omega_k \tau)} (\vec{E}_j \cdot \vec\nabla)(\vec{E}_k \cdot \vec\nabla f_0).
\end{equation*}
Here, $f_{2}^{\omega_j, \omega_k}$ denotes the second-order correction to the electron distribution function driven by the pseudo-electric and in-plane electric fields.
\noindent For strain driven by a piezoelectric voltage, the resulting pseudo-electric field can be expressed as (see Supplementary Information Section~VI),
\begin{equation}\label{eqn:E_pseudo}
    \vec E_{\omega_m}=\chi \frac{\partial \vec A_{\omega_m}}{\partial t}=-\chi\frac{\beta}{2 a}\left(\frac{d\vec{\mathcal{U}}}{dV}\right)_{V_0} V_m \omega_m \cos(\omega_m t)
\end{equation}
where $\vec{\mathcal{U}}$ depends on symmetrized strain tensor (Supplementary Information equation~(45)). This pseudo-electric field gives rise to a transverse Hall-like response (see Supplementary Information equation~(67) for detailed expression),
\begin{align}\label{eqn:transverse_wmpmwc}
    j_{y}^{\w_m\pm\w_c}&\propto\pm\chi \V_yV_m \w_m E_0 \sin[(\w_m\pm\w_c)t].
\end{align}
Both the dynamic strain-induced anomalous Hall component $V_{xy}^{\omega_m}(I_{xx}^{\omega_c} = 0)$ and the band velocity-driven $V_{xy}^{\omega_m \pm \omega_c}$ arise due to the the pseudo-electric field. These signals vanish identically in a time-reversal symmetric system when subjected to a real in-plane electric field (see Table~2 and 4 of Supplementary Information).

Figure~\ref{fig:fig3_BC_modulation_TDBG}a illustrates the transverse response of TDBG at the mixed frequency $\omega_m + \omega_c$ (see Supplementary Information section~XIII for $\omega_m - \omega_c$ signal) as a function of the perpendicular electric field ($D/\epsilon_0$) and filling factor ($\nu$) at $1.5 \, \text{K}$.
Unlike the signals shown in Figure~\ref{fig:fig2_BCD_modulation_TDBG}a and \ref{fig:fig2_BCD_modulation_TDBG}b, which arise from Berry curvature and BCD contributions, the $ \omega_m + \omega_c $ response exhibits a distinctly different behavior.
Specifically, the sign of $ V_{xy}^{\omega_m + \omega_c} $ remains positive for fillings just below full-band occupation ($ \nu \lessapprox -4, 0, +4 $) and negative for fillings just above ($ \nu \gtrapprox -4, 0, +4 $), independent of $ D/\epsilon_0 $, indicative of a band velocity–driven mechanism, an aspect that we discuss later in this work.

In Figure~\ref{fig:fig3_BC_modulation_TDBG}b, the transverse voltage at $\omega_m + \omega_c$ is plotted against the the peak-to-peak piezoelectric excitation voltage ($V_\text{piezo}^\mathrm{\omega_m}=2V_m$), which represents the amplitude of the pseudo electric field, $E_m$, at a filling $\nu = 0.037$ near the CNP and $D/\epsilon_0 = 0.354~\text{V/nm}$ (See Extended Figure~\ref{fig:extended_fig_2} for scaling with $\omega_m$).
The observed behavior confirms the linear dependency of $V_{xy}^{\omega_m+\omega_c}$ on $E_m$. 
Figure~\ref{fig:fig3_BC_modulation_TDBG}c further demonstrates the linear scaling of the Hall voltage with the amplitude of the applied current ($I_{xx}^{\omega_c}$) at $\nu = 0.037$ and $D/\epsilon_0 = 0.354~\text{V/nm}$, in agreement with equation~(\ref{eqn:transverse_wmpmwc}).
To further improve our understanding, next we probe a non-flatband system, Bernal BLG.

\subsection*{Measurements on a Bernal Bilayer Graphene}
\begin{figure*}
    \centering
    \includegraphics[width=16cm]{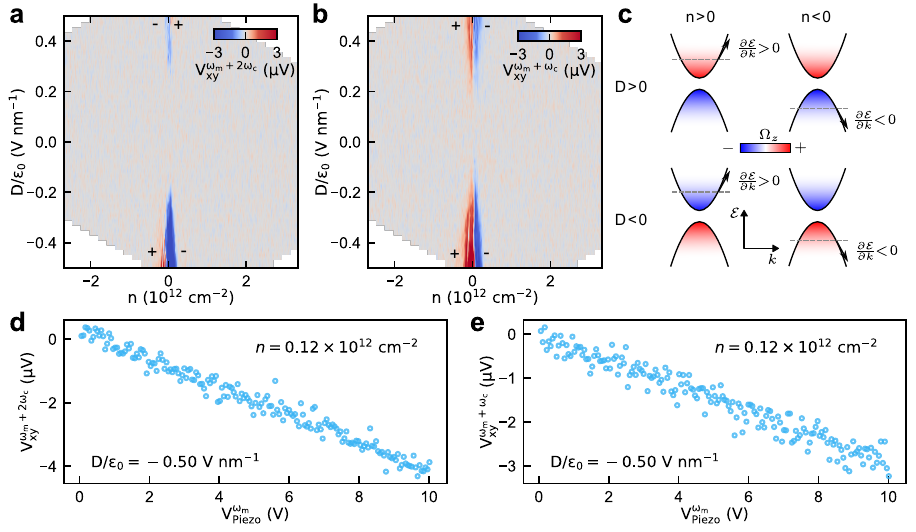}
    \caption{ \label{fig:fig4_BC_and_BCD_modulation_BLG}\textbf{Mixed-frequency Hall voltages display modulating BCD and pseudo-electric field in a complementary platform of Bernal bilayer graphene (BLG).}
\textbf{a,}~Nonlinear Hall voltage at frequency $\omega_m + 2\omega_c$ measured in AB-stacked BLG as a function of carrier density ($n$) and perpendicular electric field ($D/\epsilon_0$).
Measurements are performed at $T = 1.5 \, \text{K}$ using a current of 94 nA at 17 Hz and a 10 V peak-to-peak AC piezo excitation voltage at 177 Hz.
The signal peaks near the charge neutrality point and reverses sign with the direction of $D/\epsilon_0$, consistent with a Berry curvature dipole origin, driven by dynamic strain modulation.
\textbf{b,}~Hall-like voltage at mixed frequency $\omega_m + \omega_c$ under the same conditions. Unlike the BCD-driven signal in (a), this response does not change sign with electric field polarity, highlighting a distinct mechanism involving strain-induced pseudo-electric fields and band velocity.
\textbf{c,}~Schematic illustrating the mechanism for pseudo-electric field–driven response. The Hall-like voltage at $\omega_m + \omega_c$ arises from the combined effect of pseudo-electric field and band velocity, $ \frac{\partial\varepsilon}{\partial k}$, which changes sign in electron side ($n>0$) and hole side ($n<0$) but remains unaffected by the direction of $D/\epsilon_0$.
\textbf{d, e,}~Linear dependence of $V_{xy}^{\omega_m +2 \omega_c}$ (\textbf{d}) and $V_{xy}^{\omega_m + \omega_c}$ (\textbf{e}) with the amplitude of the AC piezo excitation voltage ($V_{\text{piezo}}^{\omega_m}$), confirms the strain-driven origin of both responses. 
}
\end{figure*}

BLG is a simpler system where similar strain-induced physics is expected. 
Strain in BLG in combination with the Berry curvature hotspots near the band edges~\cite{shimazaki_generation_2015,sui_gate-tunable_2015} induce a finite BCD giving rise to nonlinear Hall voltage near CNP~\cite{chichinadze_observation_2024}. When an oscillatory strain is introduced, BLG exhibits mixed-frequency nonlinear Hall responses analogous to those in TDBG.
Figure~\ref{fig:fig4_BC_and_BCD_modulation_BLG}a presents the Hall voltage at frequency $ \omega_m + 2\omega_c $, associated with the dynamic modulation of the BCD in BLG.
These measurements, performed at 1.5 K using an alternating current of 94 nA (17 Hz) and a 10 V peak-to-peak AC piezo excitation at 177 Hz, reveal a pronounced signal across the CNP, consistent with the formation of Berry curvature hotspots near the band edges.
Similar to ABAB-stacked TDBG, AB-stacked BLG also exhibits a sign reversal of both Berry curvature and BCD with the direction of the applied displacement field (see Supplementary Information section~II).
Accordingly, the sign of the $ \omega_m + 2\omega_c $ response reverses with field polarity, consistent with its BCD origin.

The transverse voltage at $ \omega_m + \omega_c $, arising from the interplay between the strain-induced pseudo-electric field and the applied AC current is shown in Figure~\ref{fig:fig4_BC_and_BCD_modulation_BLG}b.
Unlike the BCD-driven response in Figure~\ref{fig:fig4_BC_and_BCD_modulation_BLG}a, this signal does not reverse sign with the perpendicular electric field, consistent with a band velocity–dominated mechanism.
As schematically illustrated in Figure~\ref{fig:fig4_BC_and_BCD_modulation_BLG}c, the band velocity $ \vec{\V} = \partial \varepsilon / \partial \vec{k} $ switches sign across the CNP and does not depend on field direction.
This behavior corroborates the pseudo-electric field origin of the $ \omega_m + \omega_c $ response.
Figure~\ref{fig:fig4_BC_and_BCD_modulation_BLG}d and \ref{fig:fig4_BC_and_BCD_modulation_BLG}e further confirm the linear scaling of $V_{xy}^{\omega_m + 2\omega_c}$ and $V_{xy}^{\omega_m + \omega_c}$ with the amplitude of the AC piezo excitation voltage in alignment with the theoretical formulation~(equation~\ref{eqn:del_lambda} and \ref{eqn:transverse_wmpmwc}). 
Additional characteristics, such as the linear scaling of $V_{xy}^{\omega_m + \omega_c}$, quadratic scaling of $V_{xy}^{\omega_m + 2\omega_c}$ with AC current, and the dependence of the signals on piezo excitation frequency, are provided in the Extended Figure~\ref{fig:extended_fig_3} and \ref{fig:extended_fig_4} (see Supplementary Figure~S13, S14 and S15 for nonlinear Hall responses at other frequency components).

\section*{Discussion and conclusion}
In summary, this study introduces a novel approach to dynamically modulate Berry curvature and BCD in 2D materials using oscillatory strain via a flexible silicon nitride membrane. These results present a conclusive experimental demonstration of the dynamic modulation of the Berry curvature and BCD and pseudo-electric field’s coupling to Berry curvature. Our study under oscillatory strain uncovers the complex interplay between strain, electronic band structure, and nonlinear transport phenomena in TDBG and BLG.  The dynamic strain technique provides a significant advantage by disentangling the effects of controlled oscillatory strain from uncontrolled static strain--often introduced during fabrication--as well as from other nonlinear contributions. 

By demonstrating these effects across different 2D systems, our findings reinforce the universality of strain-induced quantum geometric and transport phenomena. While our experiments use 2D systems, the underlying concepts are very general and applicable to systems of any dimensionality as long as they possess Berry curvature. Dirac and Weyl systems~\cite{armitage_weyl_2018} would be promising candidates for future investigation.

Our findings pave a pioneering route to explore a new class of strain-engineered quantum systems, where both topological and geometric properties can be actively tuned in time. The pseudo-electric field generated by oscillatory strain possesses chirality-dependent coupling to valleys,  thereby breaking certain symmetry constraints. It can offer a novel way to probe the Chern invariant in time-reversal symmetry-protected systems. Additionally, the time-periodic modulation of quantum geometry can lead to unique light-matter interaction effects, such as valley-selective optical responses or strain-enhanced higher harmonic responses. Excitonic bands can have a finite Berry curvature~\cite{srivastava_signature_2015,zhou_berry_2015} and quantum geometry, and can be dynamically modulated as well. This can open new possibilities in the realm of strain-tunable optoelectronics and nonlinear optics. Future efforts should be directed at understanding how such temporal modulation and pseudo-electric field affect more exotic states, such as quantum Hall states~\cite{sela_quantum_2020}, superconductivity, and spontaneously symmetry-broken states.

\section*{Data Availability:}
The data that support the current study are available from the corresponding authors upon reasonable request.

\section*{Acknowledgments:}
We sincerely thank Alex Ward for providing valuable information on the piezo strain cell setup. We are grateful to Moshe Ben Shalom, Justin Song,
Raquel Queiroz and Naoto Nagaosa for insightful discussions. Darshan Joshi, Pratap Chandra Adak, and Subhajit Sinha provided constructive feedback that helped improve our manuscript. Soumyajit Samal, Rishiraj Rajkhowa, and Pratyasha Tripathy contributed significantly by assisting during device fabrication. We are thankful to Joydip Sarkar for experimental assistance. We thank Meghan P. Patankar for his assistance in making the customized PCB. M.M.D. acknowledges the Department of Atomic Energy of Government of India 12-R\&D-TFR-5.10-0100 for support, and support from J.C. Bose Fellowship JCB/2022/000045 from the Department of Science and Technology of India. M.M.D. also acknowledges support from CEFIPRA CSRP Project no. 70T07-1. U.V.W. acknowledges support from a J.C. Bose
National Fellowship. A.A. acknowledges funding from the Core Research Grant by Anusandhan National Research Foundation (ANRF, Sanction No. CRG/2023/007003), Department of Science and Technology, India, and the high performance computing (HPC) facility at IIT Kanpur including HPC 2013, and Param Sanganak. A.C. acknowledges Alexander von Humboldt foundation for financial support. R.B. acknowledges Sastra University for financial support. K.W. and T.T. acknowledge support from the Elemental Strategy Initiative conducted by the MEXT, Japan (grant no. JPMXP0112101001), and JSPS KAKENHI (grant nos. 19H05790 and JP20H00354).

\section*{Author Contributions:}
S.L. and M.A.H. fabricated the devices. S.L. did the measurements, analyzed the data and did the theoretical calculation based on Boltzmann formulation. M.A.H. and A.M. helped in measurements. M.M.D., U.V.W. and R.B. provided key initial insights for the project. A.C. and A.A. did the theoretical calculations for the consequences of static strain. L.D.V.S., S.L. and M.A.H. did the COMSOL simulation. S.L, D.A.J. and A.K. prepared the strain cell setup. A.B. fabricated the silicon nitride membranes. K.W. and T.T. grew the hBN crystals. S.L. and M.M.D. wrote the manuscript with inputs from all authors. M.M.D. initiated and supervised the project.

\section*{Competing Interests:}
The authors declare no competing interests.

\nolinenumbers
%\bibliography{bibliography}

\clearpage

\setcounter{figure}{0}
\captionsetup[figure]{labelfont={bf},labelsep=period,name={Extended Fig.}}

\begin{figure*}
    \centering
    \includegraphics[width=16.5cm]{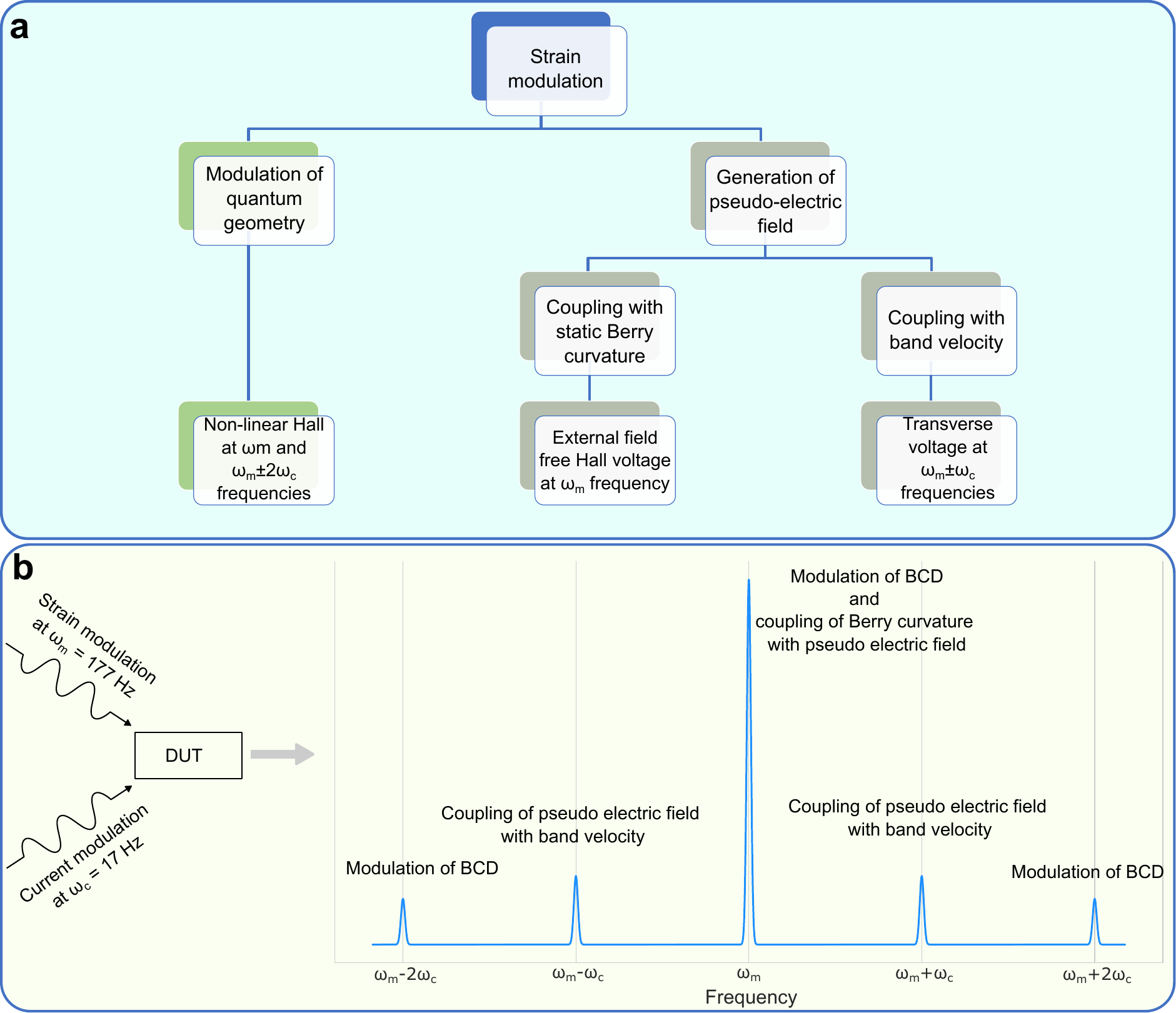}
    \caption{ \label{fig:extended_fig_0}
    \textbf{a,}~Diagram illustrating the different physical processes induced by time-modulation of strain. 
    \textbf{b,}~Schematic representation of different mixed frequency signals and their origin. 
    }
\end{figure*}

\begin{figure*}
    \centering
    \includegraphics[width=16cm]{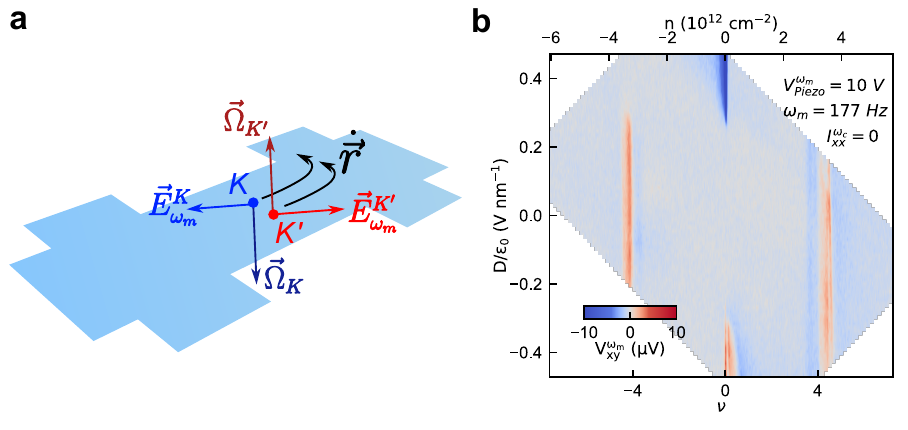}
    \caption{ \label{fig:extended_fig_1}\textbf {External electric field–free anomalous Hall effect induced by pseudo-electric field in TDBG.}
    \textbf{a,}~Schematic representation of the external field free Hall voltage at strain modulation frequency $\omega_m$. Strain modulation induces pseudo-electric field $\vec E_{\omega_m}^{K}(\vec E_{\omega_m}^{K'})$ in $K(K')$ valley. In the two valleys the pseudo-electric field as well as the Berry curvature has opposite sign. As a result anomalous velocity due to coupling of Berry curvature ($\vec\Omega$) and pseudo-electric field $-e(\vec\Omega\times \vec E_{\omega_m})$ will deflect charge carriers in both the valleys in same direction, leading to an external electric field free anomalous Hall voltage.
    \textbf{b,}~Hall voltage $V_{xy}^{\omega_m}$ measured as a function of carrier density ($n$) and perpendicular electric field ($D/\epsilon_0$) at $T = 1.5$ K, in the absence of an external current ($I_{xx}^{\omega_c} = 0$). A peak-to-peak piezoelectric modulation voltage of 10 V at frequency $\omega_m = 177$ Hz is applied to generate oscillatory strain.
    The emergence of a finite Hall voltage at the strain modulation frequency, despite the absence of applied current, demonstrates the coupling of pseudo-electric field with Berry curvature.
    }
\end{figure*}

\begin{figure*}
    \centering
    \includegraphics[width=7.5cm]{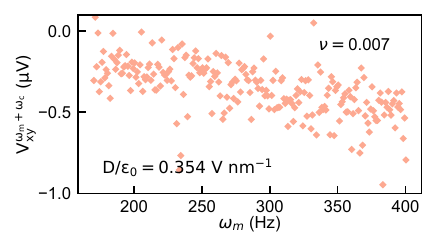}
    \caption{ \label{fig:extended_fig_2}\textbf {
    Strain modulation frequency dependence of the transverse voltage at $\omega_m + \omega_c$ in TDBG.}
    Transverse voltage $V_{xy}^{\omega_m + \omega_c}$ measured as a function of strain modulation frequency $\omega_m$ at fixed filling factor $\nu = 0.007$ and displacement field $D/\epsilon_0 = 0.354~\text{V/nm}$.
    A linear frequency dependency is observed, consistent with the expected linear scaling of the pseudo-electric field amplitude with $\omega_m$.}
\end{figure*}

\begin{figure*}
    \centering
    \includegraphics[width=16cm]{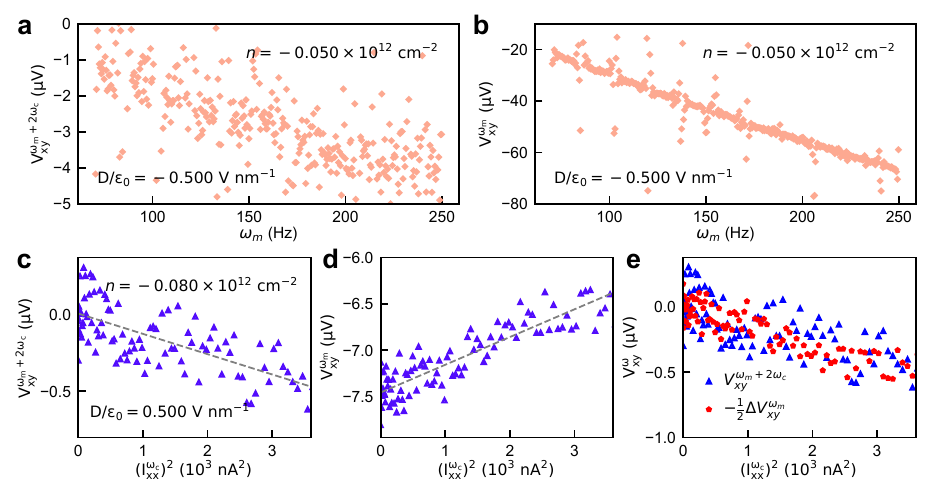}
    \caption{ \label{fig:extended_fig_3}\textbf {Strain-induced modulation of Berry curvature dipole (BCD) and pseudo-electric field in BLG.}
    These measurements are similar to the results shown in Figure~\ref{fig:fig2_BCD_modulation_TDBG}, but are performed on AB-stacked BLG.
    \textbf{a, b,} Nonlinear Hall voltages measured at mixed frequencies $\omega_m + 2\omega_c$ (a) and $\omega_m$ (b) as functions of strain modulation frequency $\omega_m$, at fixed carrier density $n = -0.050 \times 10^{12}\text{cm}^{-2}$ and displacement field $D/\epsilon_0 = -0.500~\text{V/nm}$. Both signals exhibit a clear linear frequency dependence.
    \textbf{c, d,} Nonlinear Hall voltages at $\omega_m + 2\omega_c$ (c) and $\omega_m$ (d) as functions of the square of the current, $(I_{xx}^{\omega_c})^2$, measured at $n = -0.080 \times 10^{12}\text{cm}^{-2}$ and $D/\epsilon_0 = 0.500~\text{V/nm}$. The observed quadratic dependence confirms a second-order nonlinear Hall effect in current.
    \textbf{e,} Comparison of $V_{xy}^{\omega_m + 2\omega_c}$ (blue triangles) with $-\frac{1}{2} \Delta V_{xy}^{\omega_m}$ (red circles), where $\Delta V_{xy}^{\omega_m} = V_{xy}^{\omega_m} - V_{xy}^{\omega_m}(I_{xx}^{\omega_c} = 0)$ represents the Hall voltage component driven by strain-induced BCD modulation, after subtracting the pseudo-electric-field–induced offset due to field free anomalous Hall (see Extended Figure~\ref{fig:extended_fig_1}). The close agreement between the two traces supports the BCD-modulated origin of the nonlinear Hall response.
    }
\end{figure*}

\begin{figure*}
    \centering
    \includegraphics[width=16cm]{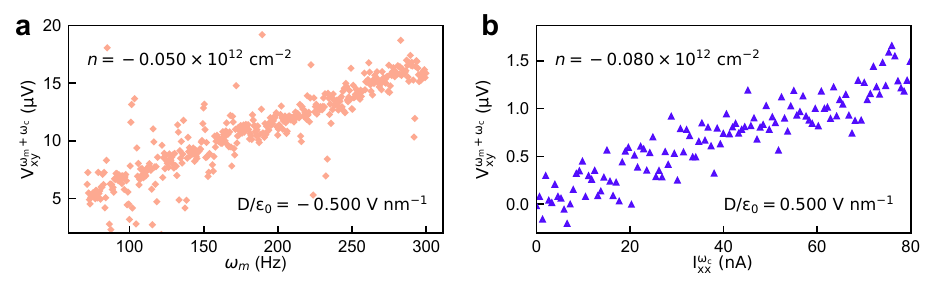}
    \caption{ \label{fig:extended_fig_4}\textbf {Transverse response at $\omega_m + \omega_c$ in BLG.}
    \textbf{a,} Transverse voltage $V_{xy}^{\omega_m + \omega_c}$ as a function of strain modulation frequency $\omega_m$, measured at carrier density $n = -0.050 \times 10^{12}\text{cm}^{-2}$ and displacement field $D/\epsilon_0 = -0.500~\text{V/nm}$. The linear scaling confirms the dynamic origin of the response consistent with a pseudo-electric field mechanism.
    \textbf{b,} $V_{xy}^{\omega_m + \omega_c}$ as a function of applied AC current amplitude $I_{xx}^{\omega_c}$ at $n = -0.080 \times 10^{12}\text{cm}^{-2}$ and $D/\epsilon_0 = 0.500~\text{V/nm}$, showing linear dependence, indicative of a first-order effect in current.
    }
\end{figure*}

\end{document}